\documentclass[10pt,twocolumn]{IEEEtran}
\usepackage{amsmath,amssymb,times}
\usepackage{epsfig}
\begin{document}

\title{\huge Study of Robust Distributed Beamforming Based on Cross-Correlation and Subspace Projection Techniques}

\author{Hang Ruan * ~and Rodrigo C. de Lamare*$^\#$ \\
$*$ Department of Electronics, The University of York, England, YO10 5BB\\
{$^\#$CETUC, Pontifical Catholic University of Rio de Janeiro, Brazil}\\
 Emails: hr648@york.ac.uk, delamare@cetuc.puc-rio.br \vspace{-0.6em}}

\maketitle
\begin{abstract}
In this work, we present a novel robust distributed beamforming
(RDB) approach to mitigate the effects of channel errors on wireless
networks equipped with relays based on the exploitation of the
cross-correlation between the received data from the relays at the
destination and the system output. The proposed RDB method, denoted
cross-correlation and subspace projection (CCSP) RDB, considers a
total relay transmit power constraint in the system and the
objective of maximizing the output signal-to-interference-plus-noise
ratio (SINR). The relay nodes are equipped with an
amplify-and-forward (AF) protocol and we assume that the channel
state information (CSI) is imperfectly known at the relays and there
is no direct link between the sources and the destination. The CCSP
does not require any costly optimization procedure and simulations
show an excellent performance as compared to previously reported
algorithms.
\end{abstract}


\section{Introduction}

Distributed beamforming has been widely investigated in wireless
communications and sensor array signal processing in recent years
\cite{r1,r2,r3}
\cite{TDS_clarke,TDS_2,switch_int,switch_mc,smce,TongW,jpais_iet,TARMO,did,badstc,baplnc,rdb}.
Such algorithms are key for situations in which the channels between
the sources and the destination have poor quality so that devices
cannot communicate directly and employ relays that receive and
forward the signals. In \cite{r2}, relay network problems are
described as optimization problems and related transformations and
implications are provided and discussed. The work in \cite{r7}
formulates an optimization problem that maximizes the output
signal-to-interference-plus-noise ratio (SINR) under total relay
transmit power constraints, by computing the beamforming weight
vector with only local information. The work in \cite{r4} focuses on
multiple scenarios with different optimization problem formulations,
in order to optimize the beamforming weight vector and increase the
system signal-to-noise ratio (SNR), with the assumption that the
global channel state information (CSI) is perfectly known. Other
works like in \cite{r5,r6} analyze power control methods based on
channel magnitude, whereas the powers of each relay are adaptively
adjusted according to the qualities of their associated channels.

However, in most scenarios encountered, the channels observed by the
relays may lead to performance degradation because of inevitable
measurement, estimation and quantization errors in CSI \cite{r12} as
well as propagation effects. These impairments result in imperfect
CSI that can affect most distributed beamforming methods
\cite{xutsa,delamaretsp,kwak,xu&liu,delamareccm,wcccm,delamareelb,jidf,delamarecl,delamaresp,delamaretvt,jioelse,rrdoa,delamarespl07,delamare_ccmmswf,jidf_echo,delamaretvt10,delamaretvt2011ST,delamare10,fa10,lei09,ccmavf,lei10,jio_ccm,ccmavf,stap_jio,zhaocheng,zhaocheng2,arh_eusipco,arh_taes,dfjio,rdrab,locsme,okspme,dcg_conf,dcg,dce,drr_conf,dta_conf1,dta_conf2,dta_ls,dvff,drr,damdc,song,wljio,barc,jiomber,saalt},
which either fail or cannot provide satisfactory performance. In
this context, robust distributed beamforming (RDB) techniques are
hence in demand to mitigate the channel errors or uncertainties and
preserve the relay system performance. The studies in
\cite{r9,r11,r12,r25} minimize the total relay transmit power under
an overall quality of service (QoS) constraint, using either a
convex semi-definite programme (SDP) relaxation method or a convex
second-order cone programme (SOCP). The works in \cite{r9,r12}
consider the channel errors as Gaussian random vectors with known
statistical distributions between the source to the relay nodes and
the relay nodes to the destination, whereas \cite{r11} models the
channel errors with their covariance matrices as a type of matrix
perturbation. The work in \cite{r11,r14,r21} presents a robust
design, which ensures that the SNR constraint is satisfied for
imperfect CSI by adopting a worst-case design and formulates the
problem as a convex optimization problem that can be solved
efficiently.

In this work, we propose an RDB technique that achieves very high
estimation accuracy in terms of channel mismatch with reduced
computational complexity, in scenarios where the global CSI is
imperfect and local communication is unavailable. Unlike existing
RDB approaches, we aim to maximize the system output SINR subject to
a total relay transmit power constraint using an approach that
exploits the cross-correlation between the beamforming weight vector
and the system output and then projects the obtained
cross-correlation vector onto subspaces computed from the statistics
of second-order imperfect channels, namely, the cross-correlation
and subspace projection (CCSP) RDB technique. Unlike our previous
work on centralized beamforming \cite{r17}, the CCSP RDB technique
is distributed and has marked differences in the way the subspace
processing is carried out. In the CCSP RDB method, the covariance
matrices of the channel errors are modeled by a certain type of
additive matrix perturbation methods \cite{r8}, which ensures that
the covariance matrices are always positive-definite. We consider
multiple source signals and assume that there is no direct link
between them and the destination. The proposed CCSP RDB technique
shows outstanding SINR performance as compared to the existing
distributed beamforming techniques, which focus on transmit power
minimization over input SNR values.

The rest of this work is organized as follows: Section II presents
the system model. Section III devises the proposed CCSP RDB method.
Section IV illustrates and discusses the simulation results. Section
V states the conclusion.

\section{System Model}

We consider a wireless communication network consisting of $K$
signal sources (one desired signal source with the others as
interferers), $M$ distributed single-antenna relays and a
destination. We assume that that direct links are not reliable and
therefore not considered. The $M$ relays receive data transmitted by
the signal sources and then retransmit to the destination by
employing beamforming, in which a two-step amplify-and-forward (AF)
protocol is considered for cooperative communications. 


In the first step, the $k$ sources transmit the
signals to the $M$ single-antenna relays according to the model
given by
\begin{equation}
{\bf x}={\bf F}{\bf s}+{\boldsymbol \nu}, \label{eq1}
\end{equation}
where the vector ${\bf s}=[s_1, s_2, \dotsb, s_K]^T \in {\mathbb
C}^{K \times 1}$ contains signals with zero mean denoted by
$s_k=\sqrt{P_{s,k}}b_k$ for $k=1,2,\dotsb,K$, where
$E[|b_k|^2]=\sigma^2_{b_k}$, $P_{s,k}$ and $b_k$ are the transmit
power and the information symbol of the $k$th signal source,
respectively. We assume that $s_1$ is the desired signal while the
remaining source signals are treated as interferers. The matrix
${\bf F}=[{\bf f}_1, {\bf f}_2, \dotsb, {\bf f}_K] \in {\mathbb
C}^{M \times K}$ is the channel matrix between the signal sources
and the relays, ${\bf f}_k=[f_{1,k}, f_{2,k}, \dotsb, f_{M,k}]^T \in
{\mathbb C}^{M \times 1}$, $f_{m,k}$ denotes the channel between the
$m$th relay and the $k$th source ($m=1,2, \dotsb, M$, $k=1,2,\dotsb,
K$). ${\boldsymbol \nu}=[\nu_1, \nu_2, \dotsb, \nu_M]^T \in {\mathbb
C}^{M \times 1}$ is the complex Gaussian noise vector at the relays
and $\sigma_{\nu}^2$ is the noise variance at each relay ($\nu_m$
\~{} ${\mathcal CN}(0,\sigma_{\nu}^2)$, which refers to the complex
Gaussian distribution with zero mean and variance $\sigma_{\nu}^2$).
The vector ${\bf x} \in {\mathbb C}^{M \times 1}$ represents the
received data at the relays. In the second step, the relays transmit
${\bf y} \in {\mathbb C}^{M \times 1}$, which is an amplified and
phase-steered version of ${\bf x}$ that can be written as
\begin{equation}
{\bf y}={\bf W}{\bf x}, \label{eq2}
\end{equation}
where ${\bf W}={\rm diag}([w_1, w_2, \dotsb, w_M]) \in {\mathbb
C}^{M \times M}$ is a diagonal matrix whose entries denote the
beamforming weights, where ${\rm diag}(.)$ denote the diagonal entry of a matrix.
Then the signal received at the destination is given by
\begin{equation}
z={\bf g}^T{\bf y}+n, \label{eq3}
\end{equation}
where $z$ is a scalar, ${\bf g}=[g_1, g_2, \dotsb, g_M]^T \in
{\mathbb C}^{M \times 1}$ is the complex Gaussian channel vector
between the relays and the destination, $n$ ($n$ \~{} ${\mathcal
CN}(0,\sigma_n^2)$) is the noise at the destination and $z$ is the
received signal at the destination. Here we assume that the noise
samples at each relay and the destination have the same power, which
means we have $P_n=\sigma_n^2=\sigma_{\nu}^2$.

Both channel matrices ${\bf F}$ and ${\bf g}$ are modeled as
Rayleigh distributed random variables, i.e., distance based
large-scale channel propagation effects that include distance based
fading and shadowing are considered. An exponential based path loss
model is described by \cite{r16}
\begin{equation}
\gamma=\frac{\sqrt{L}}{\sqrt{d^{\rho}}}, \label{eq4}
\end{equation}
where $\gamma$ is the distance-based path loss, $L$ is the known
path loss at the destination, $d$ is the distance of interest
relative to the destination and $\rho$ is the path loss exponent,
which can vary due to different environments and is typically set
within $2$ to $5$, with a lower value representing a clear and
uncluttered environment, which has a slow attenuation and a higher
value describing a cluttered and highly attenuating environment.
Shadow fading can be described as a random variable with a
probability distribution for the case of large scale fading given by
\begin{equation}
\beta=10^{(\frac{\sigma_s{\mathcal N}(0,1)}{10})}, \label{eq5}
\end{equation}
where $\beta$ is the shadowing parameter, ${\mathcal N}(0,1)$ means
the Gaussian distribution with zero mean and unit variance,
$\sigma_s$ is the shadowing spread in dB. The shadowing spread
reflects the severity of the attenuation caused by shadowing, and is
given between $0$dB to $9$dB \cite{r16}. The channels modeled with
both path-loss and shadowing can be represented as:
\begin{equation}
{\bf F}=\gamma\beta{\bf F}_0,  \label{eq6}
\end{equation}
\begin{equation}
{\bf g}=\gamma\beta{\bf g}_0,  \label{eq7}
\end{equation}
where ${\bf F}_0$ and ${\bf g}_0$ denote the Rayleigh distributed channels without large-scale propagation effects \cite{r16}.

The received signal at the $m$th relay can be expressed as:
\begin{equation}
x_m=\sum_{k=1}^K \underbrace{\sqrt{P_{s,k}}{b_k}}_{s_k}
f_{m,k}+{\nu}_m, \label{eq8}
\end{equation}
then the transmitted signal at the $m$th relay is given by
\begin{equation}
y_m={w_m}x_m. \label{eq9}
\end{equation}
The transmit power at the $m$th relay is equivalent to $E[|y_m|^2]$
so that can be written as
$\sum_{m=1}^ME[|y_m|^2]=\sum_{m=1}^ME[|{w_m}x_m|^2]$ or in matrix
form as ${\bf w}^H{\bf D}{\bf w}$ where ${\bf D}={\rm
diag}\big(\sum_{k=1}^KP_{s,k}\sigma^2_{b_k}\big[E[|f_{1,k}|^2],
E[|f_{2,k}|^2], \dotsb, E[|f_{M,k}|^2]\big]+P_n\big)$ is a full-rank
matrix, where $(.)^H$ denotes the Hermitian transpose operator. The
signal received at the destination can be expanded by substituting
\eqref{eq8} and \eqref{eq9} in \eqref{eq3}, which yields
\begin{multline}
z=\underbrace{\sum_{m=1}^M{w_m}g_m\sqrt{P_{s,1}}f_{m,1}b_1}_{\text{desired signal}}
+\underbrace{\sum_{m=1}^M{w_m}g_m\sum_{k=2}^K\sqrt{P_{s,k}}f_{m,k}b_k}_{\text{interferers}}\\
+\underbrace{\sum_{m=1}^M{w_m}g_m\nu_m+n}_{\text{noise}}.
\label{eq10}
\end{multline}
By taking expectation of the components of \eqref{eq10}, we can
compute the desired signal power $P_{z,1}$, the interference power
$P_{z,i}$ and the noise power $P_{z,n}$ at the destination as
follows:
\begin{equation}
\begin{split}
\hspace{-0.2em}P_{z,1}&=E\Big[\sum_{m=1}^M({w_m}g_m\sqrt{P_{s,1}}f_{m,1}b_1)^2\Big] \\
&= P_{s,1}\sigma^2_{b_1}\underbrace{\sum_{m=1}^M
E\Big[w_m^*(f_{m,1}g_m)(f_{m,1}g_m)^*w_m\Big]}_{{\bf w}^H E[({\bf
f}_1 \odot {\bf g})({\bf f}_1 \odot {\bf g})^H]{\bf w}},
\label{eq11}
\end{split}
\end{equation}
\begin{equation}
\begin{split}
\hspace{-0.2em}P_{z,i}& =E\Big[\sum_{m=1}^M({w_m}g_m\sum_{k=2}^K\sqrt{P_{s,k}}f_{m,k}b_k)^2\Big] \\
&=\sum_{k=2}^KP_{s,k}\sigma^2_{b_k}\underbrace{\sum_{m=1}^M
E\Big[w_m^*(f_{m,k}g_m)(f_{m,k}g_m)^*w_m\Big]}_{{\bf w}^HE[({\bf
f}_k \odot {\bf g})({\bf f}_k \odot {\bf g})^H]{\bf w}} \label{eq12}
\end{split}
\end{equation}
\begin{equation}
\hspace{-0.2em}P_{z,n}=E\Big[\sum_{m=1}^M({w_m}g_m\nu_m+n)^2\Big]
=P_n(1+\underbrace{\sum_{m=1}^ME\Big[w_m^*g_mg_m^*w_m\Big]}_{{\bf
w}^HE[{\bf g}{\bf g}^H]{\bf w}}), \label{eq13}
\end{equation}
where $*$ denotes complex conjugation. By defining $${\bf R}_k
\triangleq P_{s,k}\sigma^2_{b_k}E[({\bf f}_k \odot {\bf g})({\bf
f}_k \odot {\bf g})^H],$$ where $\odot$ is the Schur-Hadamard
product for $k=1,2,\dotsb,K$, $${\bf Q} \triangleq P_nE[{\bf g}{\bf
g}^H],$$ and the SINR is computed as:
\begin{multline}
SINR=\frac{P_{z,1}}{P_{z,i}+P_{z,n}} =\frac{{\bf w}^H{\bf R}_1{\bf
w}}{P_n +{\bf w}^H({\bf Q}+\sum_{k=2}^K{\bf R}_k){\bf w}}.
\label{eq14}
\end{multline}
It should be noted that in \eqref{eq14}, the quantities ${\bf R}_k$,
$k=1,\dotsb,K$ and ${\bf Q}$ only consist of the second-order
statistics of the channels, which means that if the channels have no
mismatches, those quantities describe the perfect knowledge of CSI.
At this point, in order to introduce errors described by ${\bf
E}=[{\bf e}_1,\dotsb,{\bf e}_K] \in {\mathbb C}^{M \times K}$ and
${\bf e} \in {\mathbb C}^{M \times 1}$ to the channels $\hat{\bf F}$
and $\hat{\bf g}$, we have
\begin{equation}
\hat{\bf f}_k={\bf f}_k+{\bf e}_k,  k=1,2,\dotsb,K,  \label{eq15}
\end{equation}
\begin{equation}
\hat{\bf g}={\bf g}+{\bf e},  k=1,2,\dotsb,K,  \label{eq15+}
\end{equation}
where $\hat{\bf f}_k$ is the $k$th mismatched channel component of
${\bf F}$. The elements of ${\bf e}_k$ for any $k=1,\dotsb,K$ and
${\bf e}$ are assumed to be for simplicity independent and
identically distributed (i.i.d) Gaussian variables so that the
covariance matrices ${\bf R}_{{\bf e}_k}=E[{\bf e}_k{\bf e}_k^H]$
and ${\bf R}_{\bf e}=E[{\bf e}{\bf e}^H]$ are diagonal, in which
case we can directly impose the effects of the uncertainties to all
the matrices associated with ${\bf f}_k$ and ${\bf g}$ in
\eqref{eq14}. By assuming that the channel errors are uncorrelated
with the channels so that $E[{\bf e}_k \odot {\bf g}]={\bf 0}$,
$E[{\bf e} \odot {\bf f}_k]={\bf 0}$, $E[{\bf e} \odot {\bf g}]={\bf
0}$ and $E[{\bf e}_k \odot {\bf f}_k]={\bf 0}$, then we can use an
additive Frobenius norm matrix perturbation method as introduced in
\cite{r8}, thus we can have the following:
\begin{equation}
\hat{\bf R}_k={\bf R}_k+{\bf R}_{{\bf e}_k}={\bf R}_k+\epsilon||{\bf R}_k||_F{\bf I}_M, k=1,\dotsb,K, \label{eq16}
\end{equation}
\begin{equation}
\hat{\bf Q}={\bf Q}+{\bf R}_{\bf e}={\bf R}_k+\epsilon||{\bf Q}||_F{\bf I}_M, k=1,\dotsb,K, \label{eq16+}
\end{equation}
\begin{equation}
\hat{\bf D}={\bf D}+\epsilon||{\bf D}||_F{\bf I}_M, \label{eq17}
\end{equation}
where $\hat{\bf R}_k$, $\hat{\bf Q}$ and $\hat{\bf D}$ are the
matrices perturbed after the channel mismatches are taken into
account, $\epsilon$ is the perturbation parameter uniformly
distributed within $(0,{\epsilon}_{max}]$ where ${\epsilon}_{max}$
is a predefined constant which describes the mismatch level. The
matrix ${\bf I}_M$ represents the identity matrix of dimension $M$
and it is clear that $\hat{\bf R}_k$, $\hat{\bf Q}$ and $\hat{\bf
D}$ are positive definite, i.e. $\hat{\bf R}_k \succ {\bf 0}
(k=1,\dotsb,K)$, $\hat{\bf Q} \succ {\bf 0}$ and $\hat{\bf D} \succ
{\bf 0}$. According to \eqref{eq14}, the robust optimization problem
that maximizes the output SINR with a total relay transmit power
constraint is written as
\begin{equation}
\begin{aligned}
\underset{\bf w}{\rm max}~~ \frac{{\bf w}^H\hat{\bf R}_1{\bf w}}{P_n+{\bf w}^H(\hat{\bf Q}+\sum_{k=2}^K\hat{\bf R}_k){\bf w}} \\
{\rm subject} ~~ {\rm to} ~~~~ {\bf w}^H\hat{\bf D}{\bf w} \leq P_T.  \label{eq18}
\end{aligned}
\end{equation}
The optimization problem \eqref{eq18} has a similar form to the
optimization problem in \cite{r4} and hence can be solved in a
closed form using an eigen-decomposition method that only requires
quantities or parameters with known second-order statistics.

\section{Proposed CCSP RDB Algorithm}

In this section, the proposed CCSP RDB algorithm is introduced. The
algorithm is considered for a system with imperfect CSI, works
iteratively to estimate and obtain the channel statistics over
snapshots. The algorithm is based on the exploitation of
cross-correction vector between the relay received data and the
system output, as well as the construction of eigen-subspaces. By
projecting the so obtained cross-correlation vector onto the
subspaces at the relays, the channel errors can be mitigated and the
result leads to a precise estimate of the beamformers. To this end,
the sample cross-correlation vector (SCV) $\hat{\bf q}(i)$ in the
$i$th iteration can be estimated by
\begin{equation}
\hat{\bf q}(i)=\frac{1}{i}\sum\limits_{j=1}^i{\bf x}(j){z^*}(j), \label{eq19}
\end{equation}
which uses sample averages that take into account all the data
observations from snapshot one to the current snapshot, where ${\bf
x}(i)$ and ${z^*}(i)$ refer to the data observation vector in the
$i$th snapshot at the relays and the system output in the $i$th
snapshot at the destination, respectively, in the presence of
channel errors. Then, we decompose the mismatched channel matrix
$\hat{\bf F}(i)$ into $K$ components as $\hat{\bf F}(i)=[\hat{\bf
f}_1(i), \hat{\bf f}_2(i), \dotsb,\hat{\bf f}_K(i)]$ and for each of
them we construct a separate projection matrix. For the $k$th
($1\leq{k}\leq{K}$) component, we compute the covariance matrix for
$\hat{\bf f}_k(i)$ and use it as an estimate of the true channel
covariance matrix instead of the mismatched channel covariance
matrices:
\begin{equation}
\hat{\bf R}_{\hat{\bf f}_k}(i)=\frac{1}{i}\sum\limits_{j=1}^i\hat{\bf f}_k(j)\hat{\bf f}_k^H(j). \label{eq20}
\end{equation}
\begin{equation}
\hat{\bf R}_{\hat{\bf g}}(i)=\frac{1}{i}\sum\limits_{j=1}^i\hat{\bf g}(j)\hat{\bf g}^H(j). \label{eq20+}
\end{equation}
Here we take an approximation for the time-averaged estimate of the
covariance matrices so that we have ${\bf R}_{{\bf
f}_k}(i)=\frac{1}{i}\sum\limits_{j=1}^i{\bf f}_k(j){\bf
f}_k^H(j)\approx\frac{1}{i}\sum\limits_{j=1}^i\hat{\bf
f}_k(j)\hat{\bf f}_k^H(j)$ and ${\bf R}_{\bf
g}(i)=\frac{1}{i}\sum\limits_{j=1}^i{\bf g}(j){\bf
g}^H(j)\approx\frac{1}{i}\sum\limits_{j=1}^i\hat{\bf g}(j)\hat{\bf
g}^H(j)$. Then the error covariance matrices ${\bf R}_{{\bf
e}_k}(i)$ and ${\bf R}_{\bf e}(i)$ can be computed as
\begin{equation}
{\bf R}_{{\bf e}_k}(i)=\epsilon||{\bf R}_{{\bf f}_k}(i)||_F{\bf I}_M. \label{eq21}
\end{equation}
\begin{equation}
{\bf R}_{{\bf e}}(i)=\epsilon||{\bf R}_{\bf g}(i)||_F{\bf I}_M. \label{eq21+}
\end{equation}
In order to eliminate or reduce the errors ${\bf e}_k(i)$ from
$\hat{\bf f}_k(i)$ and ${\bf e}$ from $\hat{\bf g}(i)$, the SCV
obtained in \eqref{eq19} can be projected onto the subspaces
described by
\begin{equation}
{\bf P}_k(i)=[{\bf c}_{1,k}(i),\dotsb,{\bf c}_{N,k}(i)][{\bf
c}_{1,k}(i),\dotsb,{\bf c}_{N,k}(i)]^H, \label{eq22}
\end{equation}
and
\begin{equation}
{\bf P}(i)=[{\bf c}_{1}(i),\dotsb,{\bf c}_{N}(i)][{\bf
c}_{1}(i),\dotsb,{\bf c}_{N}(i)]^H, \label{eq22+}
\end{equation}
where ${\bf c}_{1,k}(i),{\bf c}_{2,k}(i),\dotsb,{\bf c}_{N,k}(i)$
and ${\bf c}_{1}(i),{\bf c}_{2}(i),\dotsb,{\bf c}_{N}(i)$ are the
$N$ principal eigenvectors of the error spectrum matrix ${\bf
C}_k(i)$ and ${\bf C}(i)$, respectively, defined by
\begin{equation}
\begin{split}
{\bf C}_k(i) & \triangleq
\int\limits_{\epsilon\rightarrow{0}^{+}}^{\epsilon_{max}}E[\hat{\bf
f}_k(i)\hat{\bf f}_k^H(i)]d\epsilon
\vspace{-0.5em}\label{eq23}
\end{split}
\end{equation}
and
\begin{equation}
\begin{split}
\vspace{-0.5em}{\bf C}(i) & \triangleq \int\limits_{\epsilon\rightarrow{0}^{+}}^{\epsilon_{max}}E[\hat{\bf g}(i)\hat{\bf g}^H(i)]d \epsilon. 
\label{eq23+}
\end{split}
\end{equation}
Since we have already assumed that ${\bf e}_k(i)$ and ${\bf e}(i)$
are uncorrelated with ${\bf f}_k(i)$ and ${\bf g}(i)$, if $\epsilon$
follows a uniform distribution over the sector $(0,\epsilon_{max}]$,
by approximating $E[{\bf f}_k(i){\bf f}_k^H(i)]\approx{\bf R}_{{\bf
f}_k}(i)$, $E[{\bf e}_k(i){\bf e}_k^H(i)]\approx{\bf R}_{{\bf
e}_k}(i)$, $E[{\bf g}(i){\bf g}^H(i)]\approx{\bf R}_{\bf g}(i)$ and
$E[{\bf e}(i){\bf e}^H(i)]\approx{\bf R}_{\bf e}(i)$, \eqref{eq23}
and \eqref{eq23+} can be simplified as
\begin{equation}
\begin{split}
{\bf C}_k(i) 
& = \epsilon_{max}{\bf R}_{{\bf f}_k}(i) +
\frac{\epsilon_{max}^2}{2}||{\bf R}_{{\bf f}_k}(i)||_F{\bf I}_M,
\vspace{-0.5em} \label{eq24}
\end{split}
\end{equation}
and
\begin{equation}
\begin{split}
\vspace{-0.5em}{\bf C}(i) 
& =\epsilon_{max}{\bf R}_{\bf g}(i) +
\frac{\epsilon_{max}^2}{2}||{\bf R}_{\bf g}(i)||_F{\bf I}_M,
\label{eq24+}
\end{split}
\end{equation}
Then the mismatched channel components are then estimated by
\begin{equation}
\hat{\bf f}_k(i)=\frac{{\bf P}_k(i)\hat{\bf q}(i)}{{\lVert{{\bf P}_k(i)\hat{\bf q}(i)}\rVert}_2}, \label{eq25}
\end{equation}
\begin{equation}
\hat{\bf g}(i)=\frac{{\bf P}(i)\hat{\bf q}(i)}{{\lVert{{\bf P}(i)\hat{\bf q}(i)}\rVert}_2}. \label{eq25+}
\end{equation}
To this point, all the $K$ channel components of $\hat{\bf f}_k(i)$
are obtained so that we have $\hat{\bf F}_k(i)=[\hat{\bf f}_1(i),
\hat{\bf f}_2(i), \dotsb, \hat{\bf f}_K(i)]$. In the next step, we use
the so obtained channel components to provide estimates for the
matrix quantities $\hat{\bf R}_k(i)$ ($k=1,\dotsb,K$), $\hat{\bf Q}(i)$ and $\hat{\bf D}(i)$
in \eqref{eq18} as follows:
\begin{equation}
\hat{\bf R}_k(i)=P_{s,k}E[(\hat{\bf f}_k(i) \odot \hat{\bf g}(i))(\hat{\bf f}_k(i) \odot \hat{\bf g}(i))^H], \label{eq26}
\end{equation}
\begin{equation}
\hat{\bf Q}(i)=P_nE[\hat{\bf g}(i)\hat{\bf g}^H(i)], \label{eq26+}
\end{equation}
\begin{equation}
\hat{\bf D}(i)={\rm diag}\Big(\sum_{k=1}^KP_{s,k}[E[|\hat{f}_{1,k}(i)|^2], \dotsb, E[\hat{f}_{M,k}(i)|^2]]+P_n\Big). \label{eq27}
\end{equation}
To proceed further, we define $\hat{\bf U}(i)=\hat{\bf Q}(i)+\sum_{k=2}^K\hat{\bf R}_k(i)$ so that \eqref{eq18} can be written as
\begin{equation}
\begin{aligned}
\underset{{\bf w}(i)}{\rm max}~~ \frac{{\bf w}^H(i)\hat{\bf R}_1(i){\bf w}(i)}{P_n+{\bf w}^H(i)\hat{\bf U}(i){\bf w}(i)} \\
{\rm subject} ~~ {\rm to} ~~~~ {\bf w}^H(i)\hat{\bf D}(i){\bf w}(i)
\leq P_T.  \label{eq28}
\end{aligned}
\end{equation}
To solve the optimization problem in \eqref{eq28}, the weight vector is
rewritten as
\begin{equation}
{\bf w}(i)=\sqrt{p}{\bf D}^{-1/2}(i)\tilde{\bf w}(i), \label{eq29}
\end{equation}
where $\tilde{\bf w}(i)$ satisfies $\tilde{\bf w}^H(i)\tilde{\bf w}(i)=1$. Then \eqref{eq28} can be rewritten as
\begin{equation}
\begin{aligned}
\underset{p,\tilde{\bf w}(i)}{\rm max}~~\frac{p\tilde{\bf w}^H(i)\tilde{\bf R}_1(i)\tilde{\bf w}(i)}{p\tilde{\bf w}^H(i)\tilde{\bf U}(i)\tilde{\bf w}(i)+P_n} \\
{\rm subject} ~~ {\rm to} ~~~~ ||\tilde{\bf w}(i)||^2=1, p \leq P_T, \label{eq30}
\end{aligned}
\end{equation}
where $\tilde{\bf R}_1(i)=\hat{\bf D}^{-1/2}(i)\hat{\bf R}_1(i){\bf
D}^{-1/2}(i)$ and $\tilde{\bf U}(i)=\hat{\bf D}^{-1/2}(i)\hat{\bf U}(i)\hat{\bf
D}^{-1/2}(i)$. As the objective function in \eqref{eq30} increases
monotonically with $p$ regardless of $\tilde{\bf w}(i)$, which means
the objective function is maximized when $p=P_T$, hence \eqref{eq30}
can be simplified to
\begin{equation}
\begin{aligned}
\underset{\tilde{\bf w}(i)}{\rm max}~~\frac{P_T\tilde{\bf w}^H(i)\tilde{\bf R}_1(i)\tilde{\bf w}(i)}{P_T\tilde{\bf w}^H(i)\tilde{\bf U}(i)\tilde{\bf w}(i)+P_n} \\
{\rm subject} ~~ {\rm to} ~~~~ ||\tilde{\bf w}(i)||^2=1, \label{eq31}
\end{aligned}
\end{equation}
or equivalently as
\begin{equation}
\begin{aligned}
\underset{\tilde{\bf w}(i)}{\rm max}~~\frac{P_T\tilde{\bf w}^H(i)\tilde{\bf R}_1(i)\tilde{\bf w}(i)}{\tilde{\bf w}^H(i)(P_n{\bf I}_M+P_T\tilde{\bf U}(i))\tilde{\bf w}(i)} \\
{\rm subject} ~~ {\rm to} ~~~~ ||\tilde{\bf w}(i)||^2=1, \label{eq32}
\end{aligned}
\end{equation}
in which the objective function is maximized when $\tilde{\bf w}(i)$ is
chosen as the principal eigenvector of $(P_n{\bf I}_M+P_T{\tilde{\bf
U(i)}})^{-1}\tilde{\bf R}_1(i)$ \cite{r4}, which leads to the solution for
the weight vector of the distributed beamformer with channel errors
given by
\begin{multline}
{\bf w}(i)=\sqrt{P_T}\hat{\bf D}^{-1/2}(i){\mathcal P}\{(P_n{\bf I}_M \\ +\hat{\bf D}^{-1/2}(i)\hat{\bf U}(i)\hat{\bf D}^{-1/2}(i))^{-1} \hat{\bf D}^{-1/2}(i)\hat{\bf R}_1(i)\hat{\bf D}^{-1/2}(i)\}, \label{eq33}
\end{multline}
where ${\mathcal P}\{.\}$ denotes the principal eigenvector
corresponding to the largest eigenvalue. Then the maximum achievable
SINR of the system in the presence of channel errors is given by
\begin{multline}
{\rm SINR}_{max}=P_T{\lambda}_{max}\{(P_n{\bf I}_M+\hat{\bf D}^{-1/2}(i)\hat{\bf U}(i)\hat{\bf D}^{-1/2}(i))^{-1} \\
\hat{\bf D}^{-1/2}(i)\hat{\bf R}_1(i)\hat{\bf D}^{-1/2}(i)\}, \label{eq34}
\end{multline}
where $\lambda_{max}$ is the maximum eigenvalue. In order to
reproduce the proposed CCSP RDB algorithm, we use
\eqref{eq19}-\eqref{eq20+}, \eqref{eq24}-\eqref{eq27}, \eqref{eq33}
and \eqref{eq34} for each iteration.


\section{Simulations}

In the simulations, we compare the proposed CCSP RDB algorithm with
several existing robust approaches \cite{r11,r12,r14,r20,r21,r25}
(i.e. worst-case SDP online programming) in the presence of
imperfect CSI. The simulation metrics considered include the system
output SINR versus input SNR, snapshots as well as the maximum
allowable total transmit power $P_T$. In some scenarios, we consider
that the interferers are strong enough as compared to the desired
signal and the noise. In all simulations, the system input SNR is
known and can be controlled by adjusting only the noise power. Both
channels ${\bf F}$ and ${\bf g}$ follow the Rayleigh distribution,
whereas the mismatch is only considered for ${\bf F}$. The shadowing
and path loss parameters employ $\rho=2$, the source-to-destination
power path loss is $L=10$dB and the shadowing spread is
$\sigma_s=3$dB. The distances of the source-to-relay links
$d_{s,r_m}$ ($m=1,\dotsb,M$) are modeled as pseudo-random in an area
defined by a range of relative distances based on the
source-to-destination distance $d_{s,d}$ which is set to $1$, so as
the source-to-relay link distances $d_{s,r_m}$ are decided by a set
of uniform random variables distributed between $0.5$ to $0.9$, with
corresponding relay-source-destination angles $\theta_{r_m,s,d}$
randomly chosen from an angular range of $-\pi/2$ to $\pi/2$.
Therefore, the relay-to-destination distances $d_{r_m,d}$ can be
calculated using the trigonometric identity given by $$
d_{r_m,d}=\sqrt{d_{s,r_m}^2+1-2d_{s,r_m}\cos\theta_{r_m,s,d}}.$$ The
total number of relays and signal sources are set to $M=8$ and
$K=3$, respectively. The system interference-to-noise ratio (INR) is
specified in each scenario and $100$ snapshots are considered. The
number of principal components is manually selected to optimize the
performance for the CCSP RDB algorithm.

\begin{figure}[!htb]
\begin{center}
\def\epsfsize#1#2{0.99\columnwidth}
\epsfbox{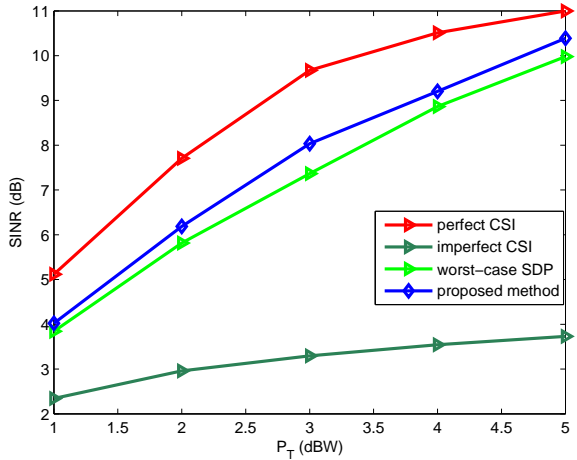} \vspace{-1em}\caption{SINR versus $P_T$,
SNR=$10$dB, $\epsilon_{max}=0.5$, INR=$10$dB} \label{figure4}
\end{center}
\end{figure}

We first examine the SINR performance versus a variation of maximum
allowable total transmit power $P_T$ (i.e. $1$dBW to $5$dBW) by
setting both SNR and INR to $10$dB. We consider that all interferers
have the same power. We also set the matrix perturbation parameter
to $\epsilon_{max}=0.5$ for all algorithms. Fig. \ref{figure4} shows
that the output SINR increases as we lift up the limit for the
maximum allowable transmit power. The proposed CCSP RDB method
outperforms the worst-case SDP algorithm and perform close to the
case with perfect CSI.

\begin{figure}[!htb]
\begin{center}
\def\epsfsize#1#2{0.99\columnwidth}
\epsfbox{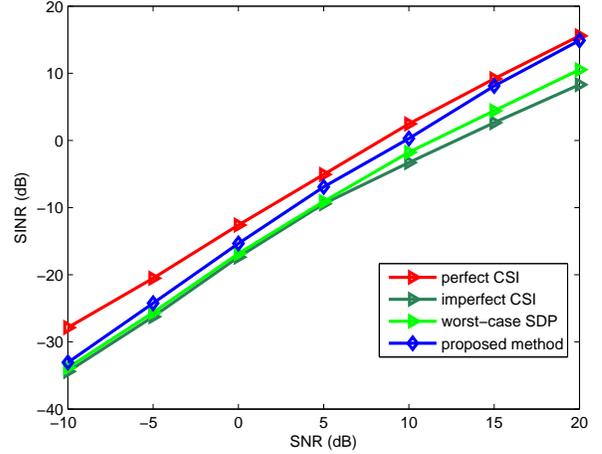} \vspace{-1em}\caption{SINR versus SNR,
$P_T=1$dBW, $\epsilon_{max}=0.2$, INR=$20$dB} \label{figure5}
\end{center}
\end{figure}

\begin{figure}[!htb]
\begin{center}
\def\epsfsize#1#2{0.99\columnwidth}
\epsfbox{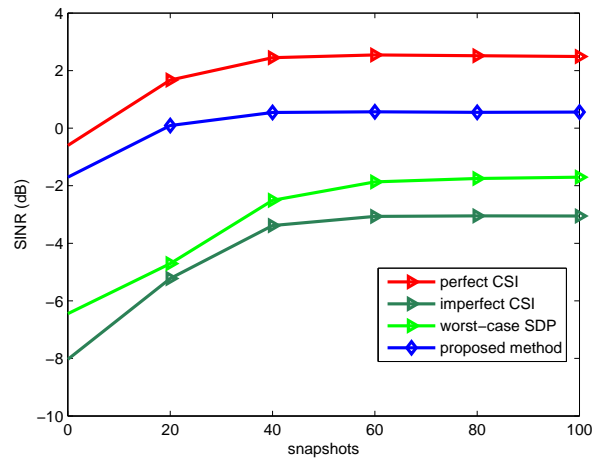} \vspace{-1em}\caption{SINR versus snapshots,
$P_T=1$dBW, $\epsilon_{max}=0.2$, SNR=$10$dB, INR=$20$dB}
\label{figure6}
\end{center}
\end{figure}

In the second example, we increase the system INR from $10$dB to
$20$dB, consider $K=3$ users but rearrange the powers of the
interferers so that one of them is much stronger than the other. We
then examine the algorithms in an incoherent scenario and set the
power ratio of the stronger interferer over the weaker to $10$. The
maximum allowable total transmit power $P_T$ and the perturbation
parameter $\epsilon_{max}$ are fixed to $1$dBW and $0.2$,
respectively. We observe the SINR performance versus SNR for these
algorithms and illustrate the results in Fig. \ref{figure5}. Then we
set the system SNR to $10$dB and observe the output SINR performance
versus snapshots as in Fig. \ref{figure6}. It can be seen that all
algorithms have performance degradation due to strong interferers as
well as their power distribution. However, the CCSP RDB algorithm
has excellent robustness in terms of SINR against the presences of
strong interferers with unbalanced power distribution.

\section{Conclusion}

We have devised the CCSP RDB approach based on the exploitation of
the cross-correlation between the received data from the relays at
the destination and the system output. The proposed CCSP RDB method
does not require any costly online optimization procedure and the
results show an excellent performance as compared to prior art.

\end{document}